\documentstyle [12pt,twoside]{article}
\begin{document}
\par\noindent{\Large\bf Invariant distributions and collisionless equilibria}
\vskip .15in
\par\noindent{\large Henry E. Kandrup}
\vskip .1in
\par\noindent{\it Department of Astronomy and Department of Physics and
Institute for Fundamental Theory, University of Florida, Gainesville, FL 
32611 USA\footnote{\small current address} and}
\vskip .1in
\par\noindent{\it Observatoire de Marseille, 2 Place Le Verrier,
F-13248 Marseille Cedex 4, France}
\vskip .3in
\small
\begin{abstract}
\noindent
This paper discusses the possibility of constructing time-independent solutions
to the collisionless Boltzmann equation which depend on quantities other than
global isolating integrals such as energy and angular momentum. The key point
is that, at least in principle, a self-consistent equilibrium can be 
constructed from {\it any} set of time-independent phase space building blocks 
which, when combined, generate the mass distribution associated with an 
assumed time-independent potential. This approach provides a way to
justify Schwarzschild's (1979) method for the numerical construction of 
self-consistent equilibria with arbitrary time-independent potentials, 
generalising thereby an approach developed by 
Vandervoort (1984) for integrable potentials. As a simple illustration, 
Schwarzschild's method is reformulated to allow for a straightforward 
computation of equilibria which depend only on one or two global integrals
and no other quantities, as is reasonable, e.g., for modeling axisymmetric 
configurations characterised by a nonintegrable potential.
\end{abstract}
\vskip .3in
\par\noindent{\bf 1 MOTIVATION}
\vskip .2in
\par\noindent
Conventional wisdom holds that galaxies in or near equilibrium can be modeled
as time-independent solutions to the collisionless Boltzmann equation. In
this context, the modeling of galaxies would seem to break logically into two 
reasonably distinct components,
namely (i) constructing time-independent solutions to the collisionless
Boltzmann equation and then (ii) determining whether said solutions are
linearly stable and otherwise viable as reasonable models of what one actually
sees. The principal focus of this paper is primarily on the former component, 
the construction of time-independent solutions, although the concluding section
will comment on issues related to viability.

Given an assumed equilibrium mass distribution ${\rho}_{0}$, and an associated 
potential ${\Phi}_{0}$ generated by the gravitational Poisson equation,
$${\nabla}^{2}{\Phi}_{0}=4{\pi}G{\rho}_{0}, \eqno(1)$$
there is a globally conserved quantity (isolating integral) $E$ reflecting 
time translational invariance. If the configuration is time-independent in an
inertial frame, this quantity is the ordinary energy $E={1\over 2}v^{2}+
{\Phi}_{0}$ (here, and henceforth, units have been chosen so 
that the stellar mass $m=1$). If instead the configuration is time-independent 
in a suitably chosen rotating frame, $E$ is the Jacobi integral.
If ${\Phi}_{0}$ manifests other continuous symmetries as well (e.g., spherical
or axial symmetry), Noether's Theorem (cf. Arnold 1989) implies that there 
will be one or more additional globally conserved quantities, say $\{I_{i}\}$. 
If there exist three independent global integrals, motion in the potential 
${\Phi}_{0}$ is integrable. If fewer than three independent global integrals
exist, the motion is nonintegrable. 

Jeans' Theorem implies that any function 
$f_{0}(E,\{I_{i}\})$ that reproduces the assumed ${\rho}_{0}$, i.e., for which 
$${\rho}_{0}=\int\,d^{3}v\,f_{0}(E,\{I_{i}\}),	\eqno(2) $$
yields a self-consistent equilibrium (cf. Binney and Tremaine 1987).
Indeed, many workers have gone further and assumed that all time-independent
equilibria must depend on such global isolating integrals. If, however, one
demands that every $f_{0}$ be a function only of some set of global isolating 
integrals,
there is an obvious critique which can be 
leveled towards numerical model-building based on Schwarzschild's (1979) 
method. This method, which involves selecting ensembles of orbit segments
that self-consistently reproduce the mass density associated with an imposed
potential, says nothing {\it a priori} about any global integrals; and, as
such, for generic potentials nothing intrinsic to the method imposes 
explicitly the demand that 
the equilibrium $f_{0}$ be a function of one or more global isolating 
integrals.

For the special case of integrable potentials, e.g., spherical configurations 
or triaxial equilibria characterised by Staeckel (1890) potentials, there 
{\it is} a direct, albeit not completely trivial, connection between 
Schwarzschild's 
method and global integrals. As discussed by Vandervoort (1984) in a slightly 
different language, if orbits are constrained by three independent global 
integrals, say 
$I_{1}({\bf r},{\bf v})$, $I_{2}({\bf r},{\bf v})$, and 
$I_{3}({\bf r},{\bf v})$, fixing the values of these integrals as $I_{1,0}$, 
$I_{2,0}$, and $I_{3,0}$ determines completely a collection of one or more
multiply 
periodic orbits in the three-dimensional configuration space, each of which 
must in principle be included in an equilibrium model with the proper relative 
weight.\footnote{Consider, for example, a spherical system with global 
integrals $E$, $J^{2}$, and $J_{z}$, and focus on motion in the equatorial
plane, i.e., $J_{x}=J_{y}=0$. Here there are an infinite number of possible 
orbits,
characterised by initial conditions corresponding to all possible points 
$(r,{\varphi})$ located in an annulus with inner and outer radii fixed by $E$ 
and $J_{z}^{2}$. The weighting implicit in eq. (3) means that all 
values of ${\varphi}$ should be treated equally, but that the relative
weighting of different $r$'s must reflect the fact that, because of 
conservation of angular momentum, orbits spend different amounts of time
at different radii. As a practical matter, however, this subtlety is arguably 
unimportant, and it may suffice computationally to consider a single orbit: If 
the radial and azimuthal periods are incommensurate, any $(x,y)$ yields an 
orbit that densely fills the annulus with the proper weight; and, even if the 
periods are not incommensurate, unless they are in a relatively low order 
resonance the orbit will generally fill a region which, to the level of
accuracy associated with one's configuration space discretisation,
is essentially dense in the annulus.}

More precisely, specifying a triplet $\{I_{1,0},I_{2,0},I_{3,0}\}$ defines a 
phase space density 
$$g_{I_{1,0},I_{2,0},I_{3,0}}({\bf r},{\bf v}){\;}
{\propto}{\;}{\delta}_{D}[I_{1}({\bf r},{\bf v})-I_{1,0}]\;
             {\delta}_{D}[I_{2}({\bf r},{\bf v})-I_{2,0}]\;
             {\delta}_{D}[I_{3}({\bf r},{\bf v})-I_{3,0}] \eqno(3)$$
and a corresponding configuration space density
$$\hskip -1.5in n_{I_{1,0},I_{2,0},I_{3,0}}({\bf r}) = \int\,d^{3}v\,
g_{I_{1,0},I_{2,0},I_{3,0}}({\bf r},{\bf v}){\;}$$
$$\hskip .9in =\int\,\int\,\int\,dI_{1}\,dI_{2}\,dI_{3}\;
{{\partial}(v_{1},v_{2},v_{3})\over 
{\partial}(I_{1},I_{2},I_{3})}\,
g_{I_{1,0},I_{2,0},I_{3,0}}({\bf r},{\bf v}), \eqno(4) $$
which, when evolved into the future using the Hamilton equations associated
with ${\Phi}_{0}$, remains unchanged. The assumption that the desired 
${\rho}_{0}$ is generated from some $f_{0}(I_{1},I_{2},I_{3})$ depending only
on global integrals means that the equilibrium distribution must be 
constructed as a superposition of solutions of the form given by eq. (3).
In this context, the proper construction of an equilibrium model using 
Schwarzschild's method thus entails three stages, namely (i) selecting all the 
orbits entering into each 
$g_{I_{1,0},I_{2,0},I_{3,0}}({\bf r},{\bf v})$ with the weights implicit in
eq. (3), (ii) performing the integration of eq. (4) to extract 
$n_{I_{1,0},I_{2,0},I_{3,0}}({\bf r})$, which incorporates the fact that each
point in configuration space is weighted in proportion to the amount of time
orbits spend there, and then (iii) choosing a superposition of 
$n_{I_{1,0},I_{2,0},I_{3,0}}({\bf r})$'s which yields the imposed
${\rho}_{0}$.

At least in the specific setting described by Vandervoort (1984), this 
interpretation breaks down for the case of nonintegrable potentials, where one 
cannot identify three global integrals, or, more generally, whenever one 
relaxes the demand that $f_{0}$ be realisable as a function of global 
isolating integrals. However, as described more carefully later on in this
paper, one can still capture the essential aspect of Vandervoort's analysis, 
namely that appropriately identified orbit segments yield the natural 
time-independent building blocks in terms of which to construct a 
self-consistent model.

Nonintegrable potentials generically admit two different classes of orbits,
namely regular and chaotic. Regular orbits in a nonintegrable potential behave 
qualitatively like orbits in an integrable potential in that they are 
multiply periodic and, even more importantly, are restricted to a three-
(or lower-)dimensional hypersurface in the six-dimensional phase space.
It follows that, even though they do not admit three global integrals, they 
must be constrained (cf. Lichtenberg and Lieberman 1992) by what are sometimes 
termed ``local integrals.''\footnote{
A good example of a local integral is the so-called third integral associated
in some cases with a nonintegrable, time-independent, axisymmetric potential,
where the only global integrals are energy $E$ and rotational angular 
momentum $J_{z}$. 
How regular orbits in a nonintegrable potential differ from orbits in an 
integrable potential is clearly stated on p. 49 of Lichtenberg and 
Lieberman (1992): ``Since the regular trajectories depend discontinuously
on initial conditions, their presence does not imply the existence of an
isolating integral (global invariant) or symmetry of the system. However,
regular trajectories, when they exist, represent exact invariants of the
motion.''}
For this reason, regular orbits in nonintegrable potentials define 
time-independent building blocks in the same sense as do orbits in integrable 
potentials. Chaotic orbits are very different in that they are intrinsically 
aperiodic and densely fill phase space regions that are necessarily higher 
than three-dimensional. However, chaotic phase space regions can still be 
characterised by invariant distributions which, when evolved into the future, 
remain unchanged; and one can use these invariant distributions as an 
additional set of building blocks when constructing self-consistent equilibria.

This alternative ``orbital'' interpretation of the building blocks for 
self-consistent equilibria is important for at least two reasons. (1) It is 
possible to generate analytically exact two-integral models for axisymmetric 
configurations which are characterised by nonintegrable potentials (cf. Hunter 
and Qian 1993), including potentials where motion in the meridional plane is 
chaotic. However, reproducing these models numerically using Schwarzschild's 
method must entail sampling a collection of time-independent building blocks 
more complex than those associated with integrable potentials. (2) Because 
global integrals are associated with continuous symmetries, one might expect 
generically that, for genuinely three-dimensional potentials, there will not 
exist any global integral aside from the energy (or the Jacobi integral). If, 
however, one demands that the equilibrium distribution be a function only of 
the energy $E$, so that $f_{0}=f_{0}(E)$, one concludes (cf. Binney and 
Tremaine 1987) that the mass distribution ${\rho}_{0}$ must be spherical. 
(This is the analogue of the well known theorem from stellar structure that 
all static perfect fluid stars are spherical.) In other words, every 
nonrotating triaxial equilibrium must depend on something other than the 
energy, either additional global integrals, as for Staeckel potentials, or
``local'' integrals, as is implicit in the construction of cuspy triaxial
models by Merritt and Fridman (1996) or Siopis (1997). 

Section 2 discusses more carefully the basic building blocks of a 
self-consistent equilibrium, allowing explicitly for the possibility of
nonintegrable potentials that admit chaotic orbits. Section 3 then illustrates
how, for the simple case of nonintegrable equilibria depending only on one
or two global isolating integrals (and nothing else), Schwarzschild's method
can be reformulated in terms of an appropriate set of time-independent 
building blocks. Section 4 concludes by describing straightforward 
generalisations to construct equilibria that do not depend simply on global 
integrals and then discussing the issue of whether such ``more complex'' 
equilibria are physically viable.
\vskip .2in
\par\noindent{\bf 2 INVARIANT DISTRIBUTIONS AND SELF-CONSISTENT EQUILIBRIA}
\vskip .1in
\par\noindent
It is often asserted glibly that ``any equilibrium solution $f_{0}$
to the collisionless Boltzmann equation must be given as a function of
the time-independent integrals of the motion associated with the potential
${\Phi}_{0}$ generated self-consistently from $f_{0}$.'' In point of fact,
however, this statement is an oversimplification and requires some 
careful thought. Should one demand, as is often assumed, at least
tacitly, that $f_{0}$ depends only on the global isolating integrals, such 
as energy $E$ (or the Jacobi integral
for a rotating configuration) and angular momentum $J_{z}$,
or can one instead allow for equilibria $f_{0}$ that depend on the 
``local'' isolating integrals which, in a generic nonintegrable 
potential, make regular orbits regular, i.e., restrict them to a 
lower-dimensional phase space hypersurface (cf. Lichtenberg and 
Lieberman 1992)? Could one, for example, try to construct models
which assign regular and chaotic orbits on the same $E$-$J_{z}$ 
hypersurface different weights, or must one sample each constant $E$-$J_{z}$
hypersurface uniformly?

Arguably, the only crucial point is that a time-independent solution 
to the collisionless Boltzmann equation must be constructed from
time-independent building blocks, so as to ensure that, if initial
data be evolved into the future along the characteristics associated
with the self-consistent potential, the form of $f_{0}$ will remain
unchanged. The {\it easiest} way to do this, both conceptually and
practically, is to demand that $f_{0}$ be given as a function of one
or more global isolating integrals, say $E$ and $I$. The obvious point
is that any $f_{0}(E,I)$ which implies the proper mass density 
${\rho}_{0}$, and hence the proper potential ${\Phi}_{0}$, will yield a
time-independent solution since, by assumption, both $E$ and $I$ are
time-independent constants of the motion. In other words, $dE/dt=dI/dt=0$
implies ${\partial}f_{0}/{\partial}t=0$. This is of course the
standard way of showing that time-independent conserved quantities can
be used to construct a self-consistent equilibrium. 

However, there is another, more ``microscopic,'' viewpoint. Specifically, 
viewed in terms of the orbits associated with the equilibrium (i.e., 
characteristics associated with the Boltzmann equation), this construction 
works because such an $f_{0}$ implies that the phase space number density 
is constant on hypersurfaces of constant $E$ and $I$. This constancy means 
that the orbit ensemble that generates $f_{0}$ must involve a uniform, i.e.,
microcanonical, sampling of each constant $E-I$ phase space hypersurface, but 
Hamilton's equations for motion in a fixed ${\Phi}_{0}$ imply that such a 
population is invariant under time translation.

It would seem clear from this latter viewpoint that choosing $f_{0}$ to be a 
function only of global isolating integrals is not necessary, at least in 
principle. {\it A priori}, a self-consistent equilibrium $f_{0}$ can be 
constructed from {\it any} collection of time-independent building blocks 
which successfully reproduces the assumed potential ${\Phi}_{0}$. The key 
point, then, as stressed, e.g., by Ott (1993), is that, assuming the validity 
of the Ergodic Theorem, for flows in a fixed time-independent potential 
{\it any} orbit is ergodic in an appropriately interpreted subspace, so that a 
microcanonical population of the appropriate subspace yields 
a time-independent building block. 

Regular orbits are multiply periodic and, as such, are characterised (in an
orbit-averaged sense) by a density that is time-independent, so that
they can be treated individually as time-independent building blocks, with
a density ${\rho}$ proportional to the time that the orbit spends in the 
neighborhood of each point.
In this sense, regular orbits in a nonintegrable potential can be exploited 
in the same way as the orbits in an integrable potential, even though it
is seemingly impossible to identify explicitly the forms of the ``local
integrals,'' and even though regularity is not attributable directly to a
continuous symmetry. 

Chaotic orbits are not periodic, so that this naive argument does not hold. 
However, it would still seem possible to identify an appropriate set of
time-independent chaotic building blocks. For a fixed value of $E$ (and
any other global integral), the constant $E$ (or $E$-$I$) phase space
hypersurface divides naturally into regular and chaotic regions. The chaotic
region divides in turn into one or more subregions which are connected in
the sense that an orbit starting in any part of a subregion will eventually
pass arbitrarily close to every other part of that subregion. The important
point then is that a uniform, i.e., microcanonical, population of any connected
chaotic domain defines a time-independent building block. Why this should 
be so is easy to understand: Time translation using Hamilton's equations 
moves each phase space point in the chaotic domain to another point in the 
same domain, but the only initial distribution invariant under time 
translation using Hamilton's equations is a constant density distribution. 
Integrating this
phase space building block over the allowed range of velocities yields a
configuration space density which, as for the regular orbits, is proportional
to the amount of time that a representative chaotic orbit in the domain
spends in the neighborhood of each point ${\bf r}$.

Assuming the validity of the Ergodic Theorem for individual connected chaotic 
domains, it is relatively simple to generate such invariant distributions 
numerically. Specifically, one knows that, when evolved into the future, a 
generic ensemble of initial conditions located anywhere in the domain will
evidence a coarse-grained approach towards the invariant microcanonical 
distribution. In fact, this has been confirmed by numerical experiments (cf.
Kandrup and Mahon 1994, Mahon, Abernathy, Bradley, and Kandrup 1995, Merritt 
and Valluri 1996) which have shown that, for chaotic flows in a variety of
different potentials, reduced distribution functions (like $f({\bf r})$ or
$f({\bf v})$) exhibit an apparent exponential in time approach towards an
invariant reduced distribution on a time scale ${\tau}$ that correlates with
the value of the largest Lyapunov exponent.

Viewed in this fashion, regular and chaotic orbits can be used to define
time-independent building blocks in exactly the same way, the only difference
being that chaotic building blocks are intrinsically higher-dimensional.
When evolved into the future, initial conditions corresponding to regular and
chaotic orbits both yield trajectories which, in an asymptotic, late 
time limit, converge towards time-independent invariant distributions. 

The crucial point in all of this is that, in principle, a library comprised 
of all possible invariant distributions, both those corresponding to individual
regular orbits and those corresponding to individual connected chaotic phase 
space regions, should constitute a complete set of building blocks in terms of 
which to construct self-consistent models of a galaxy with the specified 
potential ${\Phi}_{0}$. In the real world, one cannot construct such a 
library, which would contain an infinite number of building blocks. However, 
one {\it can} construct a large, but finite, library and then sample that 
library in an attempt to select appropriate combinations that reproduce a 
suitably discretised version of the assumed ${\Phi}_{0}$. This is the essence 
of what Schwarzschild's (1979) method can, and should, do when applied to a 
generic nonintegrable potential that admits both regular and chaotic orbits. 
It is also evident that, in principle, nothing stops one from trying to 
construct equilibria that contain only regular (or perhaps only chaotic) 
orbits, although one might imagine that it would be very hard to reproduce a 
smooth potential ${\Phi}_{0}$ with a collection of orbits that systematically 
avoids significant phase space regions or, especially, probes the phase space 
in an exceedingly irregular fashion.

For the case of a generic rotating, axisymmetric equilibrium, there 
are two global isolating integrals, namely $E$ and $J_{z}$, associated
respectively 
via Noether's Theorem with symmetries with respect to time translations
and rotations about the $z$-axis. The potential may in fact be
integrable, so that there are three global isolating integrals, but
this is not necessarily the case. In general, a rotating,
axisymmetric, time-independent potential will be nonintegrable and
admit both chaotic and regular orbits, each of which defines
time-independent building blocks. It follows that, if one allows for
local integrals, one can, at least in principle, try to construct
equilibria that do not sample constant $E$-$J_{z}$ surfaces uniformly.
One could, e.g., try to construct models which exclude all chaotic
orbits. Analytic approaches to constructing self-consistent
axisymmetric equilibria, as developed, e.g., by Hunter and Qian (1994), 
neglect this possibility altogether and focus exclusively on solutions
for which $f_{0}=f_{0}(E,J_{z})$, so that the phase space density is
constant on hypersurfaces of constant $E$ and $J_{z}$. Whether this is well
motivated physically, or whether this is only a useful analytic
simplification, is not completely clear at the present time. 

It should be stressed that great care must be taken in identifying the 
invariant distributions associated with (ensembles of) chaotic orbits.
As has long been known from numerical investigations of simple maps (cf. 
Lieberman and Lichtenberg 1972), the presence of cantori (cf. Aubry and 
Andre 1978, Mather 1982) or an Arnold (1964) web allows for the possibility
of chaotic near-invariant distributions which, albeit not strictly 
time-independent, can, at least in the absence of any perturbations, behave 
for very long times as if they were essentially time-independent 
distributions. As discussed, e.g., in Mahon, Abernathy, Bradley, and Kandrup
(1995), the crucial point here that cantori and Arnold webs serve as partial
obstructions which, although they cannot completely block motion between
different phase space regions, can significantly impede phase space transport.
It follows that, even if a single chaotic region is connected, it may appear 
partitioned into disjoint regions even over relatively long time scales.

This phenomenon is problematic. A putative self-consistent equilibrium 
generated with a  near-, rather than true, invariant distribution cannot be 
a true self-consistent equilibrium. On sufficiently long time scales, the
orbital population associated with the distribution will change, occasioning 
changes in the mass distribution, the potential, and so forth. One might 
nevertheless want to argue that, if the time scale associated with this 
phenomenon is sufficiently long, this very slow effect will be irrelevant 
astronomically, so that one can speak of nearly self-consistent equilibria 
that can exist for times much longer than the age of the Universe, $t_{H}$. 
However, this argument is probably 
wrong. Real astronomical systems involve $N$-body realisations of 
self-consistent equilibria which, heuristically, are presumed to behave like 
smooth three-dimensional Hamiltonian systems perturbed by friction and 
noise. However, numerical experiments involving perturbations of motion in a
fixed potential indicate (Habib, Kandrup, and Mahon 1996, 1997) that
even very weak friction and noise can dramatically accelerate phase space 
transport through cantori or along an Arnold web, occasioning significant 
changes in an initial near-invariant distribution on comparatively short time 
scales. Trying to estimate the longevity of a near-invariant distribution 
without allowing for the effects of perturbations is unquestionably a very 
bad idea.
\vskip .2in
\par\noindent{\bf 3 A VARIANT OF SCHWARZSCHILD'S METHOD FOR ONE- AND 
TWO-}
\vskip .05in
\par {\bf INTEGRAL DISTRIBUTIONS}
\vskip .1in
\par\noindent
The objective of this Section is to reformulate Schwarzschild's method in
terms of the natural set of building blocks so as to permit the construction 
of equilibrium models $f_{0}$ which are assumed to depend on one or two 
global integrals and to exhibit no additional dependence on any nonclassical 
local integrals. This is a straightforward generalisation of an approach 
proposed by Vandervoort (1984) for the construction of three-integral 
equilibria. 

Start by specifying a time-independent potential ${\Phi}_{0}({\bf r})$,
and hence a configuration space density ${\rho}_{0}({\bf r})$,\footnote{
Although it is the density, rather than the potential, that astronomers
are wont to specify, it is more natural conceptually to view ${\Phi}_{0}$
as the fundamental object, since it is the Hamiltonian associated with
${\Phi}_{0}$ that defines the time-invariant building blocks.}
which admits (say) two constants of the motion, $E$ and $I$, where $E$
is the particle energy (or, perhaps, the Jacobi integral) and $I$ is 
some other isolating integral, the form of which is assumed to be known 
explicitly.\footnote{Reformulating the following for 
equilibria admitting only one isolating integral
is completely straightforward. If the equilibrium admits three independent
integrals, it is integrable and can be addressed using the approach
described by Vandervoort (1984).}  By assumption, the desired equilibrium
(or equilibria) $f_{0}$ must be given exactly as a function
$f_{0}=f_{0}(E,I)$. The object, therefore, is to construct a discretised
approximation to a smooth $f_{0}$ of this form which reproduces the 
assumed ${\rho}({\bf r})$ self-consistently. This can be done in two 
stages, viz:
\begin{enumerate}
\item{}First grid $E$-$I$ space into a collection of cells and, for 
each pair $\{E_{i}$-$I_{j}\}$, write down the invariant distribution
$g_{ij}({\bf r},{\bf v})$ on the constant $E_{i}$-$I_{j}$
hypersurface. Use these $g_{ij}$'s to derive reduced configuration
space densities $n_{ij}({\bf r})$.
\item{}Then construct the desired numerical approximation to $f_{0}$ 
as a sum of contributions from the different invariant distributions
$g_{ij}$, with the relative weights of the different $g_{ij}$'s fixed 
by the requirement of self-consistency for the configuration space
density. 
\end{enumerate}
This construction proceeds without explicit reference to individual
orbits and, as such, provides no insight into the orbital building
blocks entering into the equilibrium. If this be perceived as a
serious lacuna, the natural tack numerically is to consider separately
the different constant $E$-$I$ hypersurfaces and, on each
hypersurface, to construct ensembles of orbit segments that reproduce 
self-consistently the invariant $g_{ij}$'s. 
\vskip .1in
\par\noindent
{\bf 3.1 Construction of the invariant distribution for fixed $E$ and
$I$}
\vskip .1in
\par
A uniform population of the phase space hypersurface of fixed $E_{i}$
and $I_{j}$ corresponds to an invariant distribution of the form 
$$g(E_{i},I_{j}){\;}{\equiv}{\;}g_{ij}({\bf r},{\bf v})=
{\cal K}\;{\delta}_{D}[E_{i}-E({\bf r},{\bf v})]\,
{\delta}_{D}[I_{j}-I({\bf r},{\bf v})], \eqno(5)$$
where ${\delta}_{D}$ denotes a Dirac delta, and the quantities $E$ 
and $I$ are viewed explicitly as functions of the phase space 
coordinates. The quantity ${\cal K}$ is a constant, whose value is
fixed by the normalisation
$$\int\,d^{3}x\,\int\,d^{3}v\,g_{ij}({\bf r},{\bf v})=1, \eqno(6)$$
where the integral extends over the allowed phase space regions.
In other words, the invariant distribution corresponds to a normalised
microcanonical population of the constant $E_{i}$-$I_{j}$ hypersurface. 

Given such a $g_{ij}({\bf r},{\bf v})$, it is straightforward to 
integrate over the velocity dependence to extract 
a reduced configuration space density $n_{ij}({\bf r})$. Because
$E$ and $I$ are known functions of ${\bf r}$ and ${\bf v}$, one can 
choose to view any two of the phase space coordinates, say $v_{y}$ 
and $v_{z}$, as functions of $E$, $I$, and the remaining four phase 
space coordinates.\footnote{ The choice of Cartesian
coordinates, implicit in the following, is only for specificity: as
far as this algorithm is concerned, the coordinate system is
completely irrelevant.} However, the Dirac deltas in eq.~(5) make
$dv_{y}$ and $dv_{z}$ integrations trivial, so that one can
immediately write down analytically a reduced
$${\tilde g}_{ij}(x,y,z,v_{x}){\;}{\equiv}{\;}{\cal K}\,\int dv_{y}\,dv_{z}
\;{\delta}_{D}[E_{i}-E({\bf r},{\bf v})]\;
{\delta}_{D}[I_{j}-I({\bf r},{\bf v})]. \eqno(7)$$
It follows that the configuration space density,
$$n_{ij}(x,y,z)=\int dv_{x}\,{\tilde g}_{ij}(x,y,z,v_{x}), \eqno(8) $$
associated with each constant $\{E_{i},I_{j}\}$ pair 
is given as a simple quadrature.  

In general, it may be impossible to perform the integral in eq.~(8)
analytically. This, however, is not a serious difficulty. Even if
known analytically, the $n_{ij}$'s must eventually be approximated by
a set of values on a configuration space grid so as to facilitate a 
comparison between the imposed density ${\rho}_{0}({\bf r})$ and the
inferred density $n({\bf r})$ constructed from the invariant
$n_{ij}$'s. 
\vskip .1in
\par\noindent
{\bf 3.2 Construction of $f_{0}(E,I)$ from the invariant distributions}
\vskip .1in
In the continuum limit, one knows that the true equilibrium
distribution 
$$f_{0}({\bf r},{\bf v})= \int\,\int\,dE\,dI\, A(E,I)
\;g_{E,I}({\bf r},{\bf v}), \eqno(9)$$
where $g_{E,I}$ is the invariant distribution for fixed $E$ and $I$,
viewed as a function of ${\bf r}$ and ${\bf v}$, and $A(E,I)$ is an
expansion coefficient, which gives the relative weights of the
different values of $E$ and $I$ entering into $f_{0}$. The discretised
construction thus involves
$$f_{0}({\bf r},{\bf v})=\sum_{i}\sum_{j} A_{ij}g_{ij}({\bf r},{\bf v}). 
\eqno(10) $$

The proper choice of weights $A_{ij}$ derives from the demand of 
self-consistency: Given $f_{0}$, one can define a density
$$n({\bf r})=\int\,d^{3}v\,f_{0}({\bf r},{\bf v})
 \eqno(11) $$
which, when discretised, becomes
$$n(x,y,z) = \sum_i\sum_j A_{ij}n_{ij}(x,y,z) \eqno(12) $$
in terms of the unknown expansion coefficients $A_{ij}$. However, demanding 
that this $n(x,y,z)$ correspond as closely as possible to the density 
$${\rho}_{0}(x,y,z)={1\over 4{\pi}G}\;{\nabla}^{2}{\Phi}_{0} \eqno(13) $$
associated with the assumed potential ${\Phi}_{0}$ then enables one to
determine the ``best'' values for the $A_{ij}$'s.
This construction is very much analogous to the ordinary
Schwarzschild method, save only that the basic building blocks are now
the reduced invariant distributions $n_{ij}$, rather than individual
orbits. 
\vskip .1in
\par\noindent
{\bf 3.3 Orbital building blocks for the invariant distributions}
\vskip.1in
One way in which to obtain insights into the orbital building blocks 
of a self-consistent model constructed using this algorithm is by 
proceeding numerically to construct 
ensembles of orbit segments which reproduce self-consistently the 
invariant distributions $g_{ij}$. In general, $g_{ij}$ will contain
contributions from both regular and chaotic orbits, each of which is
characterised separately by its own invariant distribution. 
The easiest way to construct $g_{ij}$ is probably to (1) obtain an 
invariant distribution for the chaotic orbits and then (2) augment
this by another (sub)distribution comprised of segments of regular
orbits, the latter so chosen that the combination of regular and 
chaotic orbits yields a satisfactory approximation to the true invariant 
distribution. 

The invariant distribution is approximated numerically
by binning the six-dimen-sional phase space into a collection of
six-dimensional hypercubes, and then assigning occupation numbers to
the different hypercubes which are proportional to the time that
orbits sampling the true invariant distribution reside in each cell. 
(This is justified by the Ergodic Theorem [cf. Lichtenberg and
Lieberman 1992].) In the continuum limit, the invariant distribution
corresponds to a uniform population on a four-dimensional phase space
hypersurface. Given a discretisation of the phase space coordinates, 
the invariant distribution corresponds instead to a four-dimensional 
shell in the six-dimensional phase space.

The invariant (sub-)distribution of chaotic orbits is especially easy to 
compute if, as is often the case, for fixed $E_{i}$ and $I_{j}$ the
entire chaotic region is connected in the sense that, eventually, 
every chaotic orbit will pass arbitrarily close to every point in the 
chaotic region. (For simplicity, ignore the tiny measure of chaotic orbits
trapped permanently inside invariant {\it KAM} tori.) All that one
need do is specify a (more or less arbitrary) ensemble of initial 
conditions, each corresponding to a chaotic orbit, evolve each initial
condition into the future, and wait until the evolved ensemble
approaches an invariant distribution, i.e., a uniform sampling of the
chaotic portions of the $E_{i}$-$I_{j}$ hypersurface (cf. Kandrup and 
Mahon 1994, Mahon, Abernathy, Bradley, and Kandrup 1995). 

To expedite the calculation, it is useful to evolve the initial 
conditions in the presence of very weak amplitude friction and noise, 
sufficiently weak that the values of $E$ and $I$ are almost constant 
(cf. Habib, Kandrup, and Mahon 1996, 1997). The advantage of introducing
weak friction and noise is that such small perturbations can
dramatically accelerate the overall approach towards a true invariant 
distribution by facilitating extrinsic diffusion through cantori 
and/or along an Arnold web (cf. Lichtenberg and Lieberman 1992). If 
one does not either (a) integrate for a very long time and/or (b) 
allow for such perturbing influences, one faces the problem that the 
initial ensemble may evolve towards a near-invariant distribution 
which, albeit not strictly time-independent, only changes
significantly on a very long time scale.

The contribution of different regular orbits to the invariant distribution 
can be generated using an analogue of the original Schwarzschild method.
Specify a large number of regular initial conditions and integrate each
into the future to generate a library of regular orbits. Then use a 
linear programming algorithm, or some variant thereof,  to select a 
weighted ensemble of regular orbits which, when combined with the chaotic
(sub)distribution, yields a satisfactory approximation to the true 
invariant distribution on the constant $E_{i}$-$I_{j}$ hypersurface. 

This construction of invariant distributions $g_{ij}$, and the
corresponding densities $n_{ij}$, is admittedly tedious numerically 
(albeit presumably straightforward), since it involves repeating
Schwarzschild's method for each pair $\{E_{i},I_{j}\}$. However, it is 
arguably a crucial step in obtaining a proper understanding of the
orbital structure associated with the self-consistent model since, 
as discussed already, one knows that the $g_{ij}$'s are the proper 
building blocks in terms of which to construct an equilibrium
$f_{0}(E,I)$.
\vskip .1in
\par\noindent
{\bf 3.4 A simple two-dimensional example}
\vskip .1in
\par\noindent
Consider as a pedagogical example the case of two-dimensional gravity, this
corresponding physically to a collection of infinite rods aligned along
the $z$-axis, and suppose that the configuration is rotating uniformly 
about the $z$-axis with angular velocity ${\Omega}$. It then follows that,
in the rotating frame, the configuration is characterised by a potential
${\Psi}(x,y)$ and a surface density ${\sigma}(x,y)$ related by 
$${\nabla}^{2}{\Psi}(x,y)=4{\pi}G{\sigma}(x,y) . \eqno(15) $$

Suppose then that there is only one global isolating integral, namely the 
Jacobi integral $E$, which, in terms of phase space coordinates defined
in the rotating frame, takes the form
$$E={1\over 2}v_{x}^{2}+{1\over 2}v_{y}^{2}+
{\Psi}(x,y)-{1\over 2}{\Omega}^{2}(x^{2}+y^{2}){\;}{\equiv}{\;}
{1\over 2}v_{x}^{2}+{1\over 2}v_{y}^{2}+{\Psi}_{eff}(x,y) . \eqno(16) $$
To the extent that one demands that any equilibrium $f_{0}$ associated
with this mass distribution be a function only of the isolating integral
$E$, the fundamental building block is the microcanonical phase space 
density on a constant Jacobi integral hypersurface, which, for any $E_{i}$, 
takes the form
$$g(E_{i})=g_{i}({\bf r},{\bf v})={\cal K}\,
{\delta}_{D}[E_{i}-E({\bf r},{\bf v})] . \eqno(17) $$
As will be evident from below, the normalisation constant ${\cal K}$
can be written as
$${\cal K}={1\over 2{\pi}V(E_{i})}  \eqno(18) $$
in terms of $V(E_{i})$, the area of the configuration space region
with ${\Psi}_{eff}{\;}{\le}{\;}E_{i}$. 
The reduced configuration space density $n_{i}$ associated with 
this $g_{i}$ satisfies
$$n_{i}(x,y)={\cal K}\,\int\,\int\,dv_{x}\,dv_{y}\;
{\delta}_{D}[E_{i}-E({\bf r},{\bf v})] , \eqno(19) $$
where the integrals extend over the values of $v_{x}$ and $v_{y}$ that
are allowed energetically, i.e., for which $E_{i}{\;}{\ge}{\;}{\Psi}_{eff}$.  
The $dv_{y}$ integration can be performed trivially by implementing 
the Dirac delta, allowing for nonzero contributions at two values, namely
$v_{y}={\pm}\sqrt{2(E_{i}-{\Psi}_{eff})-v_{x}^{2}}{\;}{\equiv}{\;}$ ${\alpha}$.
It follows that, for those regions in configuration space for which
${\Psi}_{eff}(x,y){\;}{\le}{\;}E_{i}$,
$$n(x,y)=2{\cal K}\,\int_{-\alpha}^{\alpha} \;
{dv_{x}\over \sqrt{{\alpha}^{2}-v_{x}^{2}}}\;. \eqno(20) $$
The remaining integral can then be performed trivially, leading to a
reduced configuration space density on the constant $E_{i}$
hypersurface of the form
$$n_{i}(x,y)=2{\pi}{\cal K}\,{\Theta}[E_{i}-{\Psi}_{eff}(x,y)] 
={1\over V(E_{i})}\,{\Theta}[E_{i}-{\Psi}_{eff}(x,y)] , \eqno(21) $$
where ${\Theta}(z)=1$ for $z{\;}{\ge}{\;}0$ and ${\Theta}=0$
otherwise. 

It follows from eq.~(21) that, independent of the specific form of the
potential ${\Psi}(x,y)$, the total configuration space surface
density, given as a sum of contributions on different constant Jacobi 
integral hypersurfaces, must be of the form 
$$n(x,y)=\sum_{i}\,A_{i}{1\over V(E_{i})}\;
{\Theta}[E_{i}-{\Psi}_{eff}(x,y)] . \eqno(22) $$
where the $A_{i}$'s give the relative weights of the different
$E_{i}$ hypersurfaces. The demand that the $n(x,y)$ of 
eq.~(22) agree as closely as possible with the ${\sigma}(x,y)$ associated with 
${\Psi}$ may then be used to identify the ``best'' values of the $A_{i}$'s.
\vskip .2in 
\par\noindent{\bf 4 DISCUSSION} 
\vskip .1in
\par\noindent
There are a number of different ways in which the algorithm described in the
preceding Section can be generalised to permit the construction of more complex
equilibria, which do not depend simply on the global integrals $E$ and $I$.
For fixed values of $E$ and $I$, it is straightforward to locate the general
locations of (at least the large) chaotic regions and, by evolving arbitrary
ensembles of initial conditions located in these regions into the future, it
is easy to derive a numerical approximation to the invariant distribution
associated with each of these chaotic regions. Given these invariant
distributions, one can then integrate over velocities to extract the chaotic
contribution $n^{c}_{ij}(x,y,z)$ to total density $n_{ij}(x,y,z)$ associated
with any pair $E_{i}$ and $I_{j}$. Subtracting $n^{c}_{ij}$ from the full
$n_{ij}$ then yields the regular contribution $n^{r}_{ij}(x,y,z)$ to the
density. However, given a knowledge of $n^{r}_{ij}$ and $n^{c}_{ij}$ 
separately, one can then attempt to construct models which assign different
relative weights to the regular and chaotic portions of the $E_{i}$-$I_{j}$
hypersurface, thus allowing one to test the prejudice of some workers that
self-consistent models should contain few, if any, chaotic orbits.

Similarly, one can identify those portions of the $E_{i}$-$I_{j}$ hypersurface
that correspond to different types of regular orbits, e.g., boxes and tubes,
and compute their relative densities, say $n^{b}_{ij}$ and $n^{t}_{ij}$,
which can in turn be used as separate building blocks. In particular, given
such a collection one can try to construct models which associate different
relative weights to boxy and/or tuby and/or chaotic orbits, and, to the extent 
that such models can be constructed, one can investigate whether the different 
phase space 
densities $f_{0}$ have obvious observational signatures which could be 
compared with real astronomical data. Is there, e.g., some natural signature
which, when observed in real galaxies, can be interpreted as evidence that 
$f_{0}$ contains a significant measure of chaotic orbits?

In principle, one can continue this process of refinement more or less
{\it ad infinitum}, identifying increasing numbers of time-independent 
building blocks associated with different regular orbits, although one's
freedom to deal with chaotic orbits is limited by the fact that there is
only one natural notion of a time-independent invariant distribution. However, 
it is not clear
that such a procedure is well motivated physically. At least heuristically,
it would seem that building an equilibrium by ``picking and choosing''
amongst individual orbits in a strongly nonintegrable potential with 
different values of local isolating integrals is akin to selecting orbits 
in an integrable potential which yield a distribution function that is a 
highly irregular function of the $I_{i}$'s. This latter procedure might 
strike one as contrived and, in any event, one knows that, in many cases, 
such irregular $f_{0}$'s are linearly unstable. Thus, e.g., it is well
known that, for a spherical equilibrium with $f_{0}=f_{0}(E,J^{2})$, 
stability or lack thereof often correlates with the sign of the partial 
derivatives ${\partial}f_{0}/{\partial}E$ and/or 
${\partial}f_{0}/{\partial}J^{2}$. In particular, population
inversions can trigger instabilities. 

If any discrete construction based on Schwarzschild's method is to be 
reasonable, there must be a sense in which, as the discretisation of the 
density becomes more refined and as the number of building blocks becomes 
larger, the solution constructed numerically converges towards a continuous 
self-consistent equilibrium. However, identifying the precise sense in which
this is so would most likely be {\it very} difficult. Mathematically, 
establishing such a convergence would probably involve a study of sequences 
of discrete Banach spaces, along the lines that have been used to study the 
convergence of finite difference schemes for solving partial differential 
equations. In that setting, a good deal is known about linear differential
equations but, if one incorporates nonlinearities and/or allows 
for an integro-differential equation -- recall that the collisionless
Boltzmann is a quadratically nonlinear integro-differential equation
-- things become much harder!

It is evident, both intuitively and from painful experience (cf. Siopis,
Athanassoula, and Kandrup 1997), that it is 
easier to approximate comparatively smooth quantities on a finite
lattice than quantities that manifest intricate structures on a variety of
different scales. For this reason, one might expect that it is much
easier to construct a satisfactory numerical representation of an
$f_{0}$ that is a function only of smoothly varying global isolating
integrals than an $f_{0}$ that depends sensitively on ``local''
integrals that manifest the details of the complex phase space
structure associated with a generic nonintegrable potential. Moreover,
even if one allows for local integrals, the numerical construction
should be more straightforward if, for example, on a constant energy
hypersurface, the phase space population is reasonably smooth, e.g.,
perhaps avoiding chaotic regions but weighting different regular regions in a 
fashion that varies smoothly with their phase space location.

Suppose that there is in fact a ``true'' $f_{0}$ involving local
integrals, generated (in principle) as an exact time-independent 
solution to the 
collisionless Boltzmann equation, to which one has constructed a 
latticized $f_{0}$ via some analogue of Schwarzschild's method, and 
that this latticized $f_{0}$ has been used to generate an ensemble of 
initial conditions to populate an $N$-body realisation of the model. 
There are then two potentially serious sources
of error: (1) The latticized approximation to $f_{0}$ could miss
important microscopic structures associated with local
integrals.
 If the ``true'' $f_{0}$ is a function only of smoothly varying
integrals like the energy $E$, allowing for as few as $20$ different energies,
as did Schwarzschild (1979), Merritt and Fridman (1996), and Siopis
(1997), may be adequate to capture the
essence of the analytic model. If, however, $f_{0}$ involves a complex
combination of local integrals as well as the energy, allowing for
$20$ values may not be enough. 
(2) Even if most/all the important microscopic structures are
adequately represented in the discretised model, the $N$-body
realisation could fail to sample them adequately. Even for very large
particle number, $N{\;}{\sim}{\;}10^{6}$ or more, there is no 
guarantee that a complex phase space will be adequately sampled. 

If, however, it is difficult for galactic astronomers to construct
$N$-body realisations of ``complex'' equilibria $f_{0}$ that involve
local integrals in a highly irregular way, nature too may find it hard.
If $N{\;}{\sim}{\;}10^{6}$ is not enough to generate a ``fair'' sampling, 
$N{\;}{\sim}{\;}10^{11}$ may also be inadequate to probe the complex 
phase space associated with a smooth potential generated as a Boltzmann 
equilibrium. Both statistical fluctuations, which will obviously be
present, and small non-Hamiltonian irregularities, which can be important
in complex Hamiltonian systems (cf. Lichtenberg and Lieberman 1992) may
tend to ``smooth out'' a complex would-be equilibrium into something
substantially simpler.

\vskip .2in
\par\noindent
{\bf ACKNOWLEDGMENTS}
\vskip .1in
\par\noindent
Work on this paper began while I was a visitor at the Observatoire de 
Marseille, where I was supported by the {\it C.N.R.S}. Additional support
was provided by the National Science Foundation through PHY92-03333. 
Portions of this paper were written while I was a visitor at the Aspen
Center for Physics.
I am grateful to Evangelia Athanassoula, Chris Hunter, Christos Siopis,
and Haywood Smith for useful comments and interactions.
\vfill\eject
\par\noindent
Arnold, V. I. 1964, Russ. Math. Surveys 18, 85.
\par\noindent
Arnold, V. I. 1989, Mathematical Methods of Classical Mechanics (Berlin:
Springer).
\par\noindent
Aubry, S. and Andre, G. 1978, in Solitons and Condensed Matter Physics,
ed. A. R. Bishop \& T. Schneider (Berlin: Springer), 264.
\par\noindent
Binney, J. and Tremaine, S. 1987, Galactic Dynamics (Princeton: Princeton
Univ. Press).
\par\noindent
Habib, S., Kandrup, H. E., \& Mahon, M. E. 1996, Phys. Rev. E 53, 5473.
\par\noindent
Habib, S., Kandrup, H. E., \& Mahon, M. E. 1997, ApJ 480, 155.
\par\noindent
Hunter, C. and Qian, E. 1993, MNRAS 262, 401.
\par\noindent
Kandrup, H. E. \& Mahon, M. E. 1994, Phys. Rev. E 49, 3735.
\par\noindent
Lichtenberg, A. J. \& Lieberman, M. A. 1992, Regular and Chaotic Dynamics
(Berlin: Springer).
\par\noindent
Lieberman, M. A. \& Lichtenberg, A. J. 1972 , Phys. Rev. A 5, 1852.
\par\noindent
Mahon, M. E., Abernathy, R. A., Bradley, B. O., and Kandrup, H. E. 1995,
MNRAS 275, 443
\par\noindent
Mather, J. N. 1982, Topology 21, 45.
\par\noindent
Merritt, D. and Fridman, T. 1996, ApJ 460, 136.
\par\noindent
Merritt, D. and Valluri, M. 1996, ApJ 471, 82.
\par\noindent
Ott, E. 1993, Chaos in Dynamical Systems (Cambridge: Cambridge Univ. Press).
\par\noindent
Schwarzschild, M. 1979, ApJ 232, 236.
\par\noindent
Siopis, C. 1997, University of Florida Ph.~D. dissertation.
\par\noindent
Siopis, C., Athanassoula, E., and Kandrup, H. E. 1997, A\&A, in preparation.
\par\noindent
Staeckel, P. 1890, Math. Ann 35, 91.
\par\noindent
Vandervoort, P. O. 1984, ApJ 287, 475.

\vfill\eject\end{document}